\documentclass[%
 reprint,
 amsmath,amssymb,
 aps,
]{revtex4-2}

\usepackage{graphicx}
\usepackage{dcolumn}
\usepackage{bm}

\usepackage{babel}

\usepackage{xcolor}

\usepackage{amsmath,amssymb}
\usepackage{siunitx}

\usepackage{array,multirow,makecell}
\usepackage{hhline}
\usepackage{ulem}

\begin{document}

\preprint{APS/123-QED}

\title{Photon Bose-Einstein condensation and lasing in semiconductor cavities}

\author{Aurelian Loirette-Pelous}
\author{Jean-Jacques Greffet}%
\affiliation{Université Paris-Saclay, Institut d'Optique Graduate School, CNRS, Laboratoire Charles Fabry, 91127, Palaiseau, France
}%




\date{\today}

\begin{abstract}
Photon Bose-Einstein condensation and photon thermalisation have been largely studied with molecular gain media in optical cavities. Their observation with semiconductors has remained elusive despite a large body of experimental results and a very well established theoretical framework. We use this theoretical framework as a convenient platform to revisit photon Bose-Einstein condensation in the driven dissipative regime and compare with the lasing regime. We discuss the thermalisation figures of merit and the different experimental procedures to asses thermalization. We compare the definitions of lasing and condensation thresholds. Finally, we explore the fluctuations of the system and their relation to the different regimes. 
\end{abstract}


\maketitle

\section{Introduction}

In 2010, experiments by Klaers et al. \cite{klaers2010thermalization,klaers2010bose} identified and demonstrated Bose-Einstein condensation of photons, a new light emission regime. {While this regime share with lasing the macroscopic occupation of one mode, cavity photons are in near-thermodynamic equilibrium.} As a direct consequence, cavity modes occupation follow a Bose-Einstein (BE) distribution and condensation is forced in the lowest energy cavity mode.

At first glance, Bose-Einstein condensation (BEC) with photons {seems} to be impossible. On the one hand, lasers are usually thought to operate far from equilibrium. On the other hand, in the so-called blackbody radiation, equilibrium between photons is reached due to walls acting as a reservoir, but the null chemical potential precludes condensation. 
{Actually, a suitable gain material such as pumped dyes molecules or semiconductors can act as a reservoir providing a photon chemical potential \cite{wurfel1982chemical}. Thermalization of the photon gas with such a reservoir is made possible with a high-Q cavity, when the number of absorption-emission cycles made by a photon before leaving the cavity becomes large.} Furthermore, the cavity introduces a band gap in the photon dispersion relation so that a lowest energy state can be defined for a given band. These ingredients are sufficient to ensure BE condensation of photons at room temperature in the weak coupling regime \cite{klaers2010bose}. 

In the last decade, the pioneering experiments \cite{klaers2010thermalization,klaers2010bose} in a dye-filled microcavity triggered a large amount of works in similar devices in order to understand further this new regime and its properties. An important issue has been to clarify the similarities and differences with the lasing regime. While the overall crossover from the standard {out-of}-equilibrium lasing phase to the BEC one has been shown to be quite smooth \cite{schmitt2015dynamics}, some features of BEC have appeared. {At equilibrium, the emission spectrum follows a BE distribution, and condensation occurs into the lowest energy cavity modes. When thermalization breaks down, major spectral alterations have been observed,} ranging from deformation of the thermal tail \cite{klaers2010thermalization} to lasing in excited modes and multimode lasing \cite{marelic2016spatiotemporal,nyman_nature_2018,rodrigues2021learning}. Early experiments investigating the second-order coherence in the BEC regime evidenced 
large fluctuations $g^{(2)}(0)=2$ even far above the condensation threshold \cite{schmitt2014PRL}. This thermal behaviour suggests a closest resemblance of a photon BEC to a pumped blackbody than to a standard laser. 
As a consequence, first order temporal coherence is also delayed to above-threshold excitation \cite{schmitt2016spontaneous}.
{In recent years, the question about the difference between BEC and lasing has been renewed due to the emergence of nanostructured cavity mirrors enabling to realize complex potentials for light \cite{dung2017variable,kurtscheid2019thermally,kurtscheid2020realizing,busley2021compressibility,walker2021bespoke}.}
Indeed, in these systems, controlling the thermalization enables, for example, the study of vortices formation and annihilation \cite{gladilin2020vortices,gladilin2021vortex,gladilin2022vortex}, or to envision analog simulation with synchronized arrays of out-of-equilibrium condensates \cite{vretenar2021modified,vretenar2021controllable,bloch2022non}. Still, in the quest for these new applications, we observe that several aspects of the problem have been overlooked so far. We list several of them in the next paragraphs.

We first note that BE condensation of photons has been observed in dye-filled microcavities \cite{klaers2010bose,nyman_nature_2018,vanoosten2018density} and plasmonic nanoparticles arrays \cite{torma2018nature}, and erbium–ytterbium co-doped fiber cavities \cite{weill2019bose}. {Alternatively, semiconductors have received much less attention up to now, in spite of being a very common and versatile active medium. In particular,  photon BEC in semiconductor-based devices is not fully recognized yet.} This is surprising in many respects. On the experimental one, {spectral} signatures hinting at thermalization and BEC of photons has been observed early in a VCSEL designed for polariton physics \cite{bajoni2007,lagoudakis2012crossover}. More recently, similar features have been observed in a commercial VCSEL \cite{barland}, suggesting that BEC (or near-equilibrium BEC) of photons could be more common than it is usually thought. A {high absorption/emission cycles number before cavity loss} has also been reported in a quantum-well photonic crystal laser \cite{takemura2019lasing}, {while not interpreted as BEC. On the theoretical side, the possibility of a chemical potential for photons has been historically demonstrated on a semiconductor example \cite{wurfel1982chemical}. Furthermore, simple and accurate models of gain and lasing in semiconductors  are available so that this system is a very good playground to explore the physics of photon Bose-Einstein condensation and lasing.

Second, finding a clear signature of photon thermalization is not an obvious task. On the experimental side, the analysis of emission spectra is often compared with a Bose-Einstein function. On the theoretical side, a dimensionless number quantifying the degree of thermalization has been introduced theoretically by some authors. A simple connection between these two approaches is still lacking.

Third, the connection between lasing and condensation is not fully understood.
While a clear threshold is observed in both cases, its exact positions differs. This may impact the interpretation of the observed phenomena. Hence, there is a need to compare the definitions of thresholds from laser and from equilibrium BE condensates physics.

Fourth, intensity correlations are often used to distinguish coherent light from stochastic light.  It is interesting to revisit condensation and lasing by studying fluctuations. While many results have been reported, the role of the degree of thermalization and the role of the $\beta$-factor of the cavity have not been fully discussed so that it is difficult to draw final conclusions. 

In this paper, we take advantage of the well-developped formalism to describe gain in semiconductors to analyse all these issues. In the next section, we present a simple unified theory of equilibrium and non-equilibrium condensation of photons in a semiconductor-based cavity. While similar to the pioneering model by Kirton and Keeling \cite{kirton_killing2013,kirton_killing_2015} for dye-filled microcavities, we show that our model provides a straightforward interpretation of the photons chemical potential. We then derive a generalized BE distribution in the driven-dissipative regime  and exhibit a dimensionless number that characterizes quantitatively the degree of thermalization. We discuss some of its properties and clarify the connection with other dimensionless numbers such as cooperativity, Knudsen number and optical thickness. In this new framework, we show how to revisit some lasing features such as gain clamping and inversion, and discuss the selection of the lasing mode. An extended definition of the equilibrium condensation threshold is also introduced for nonequilibrium systems{, and compared to the standard lasing threshold definition}. 

In Section 3, we discuss {several observables} to evaluate to which extent a device is thermalized. Equipped with the explicit form of the degree of thermalization introduced in the previous section, we can revisit the typical experimental situations. {In particular, we show that the most common practice, consisting in studying the emission spectrum, should be used with caution. }

We finally focus on the second-order coherence in Section 4. {We tackle the thermalisation issue by calculating analytically the intensity autocorrelation function $g^{(2)}(0)$ as a function of the $\beta$-factor and the degree of thermalization. }

\section{ {Equilibrium and non-equilibrium condensation of photons in a semiconductor-based cavity} }

In this section, we first summarize basic forms of the emission and absorption rate in a semiconductor. We then use this formalism to recover  the equilibrium number of photons per mode in a lossless cavity. {We finally compare this case to the one of a lossy cavity with gain operating in the so-called driven-dissipative regime}. This approach enables us to discuss in a very simple framework (i) the thermalization regime{, introducing} a degree of thermalization in a very systematic way, (ii) the connection between condensation and lasing and (iii) the definitions of their respective threshold.

\subsection{ Model of semiconductor gain medium in a cavity }\label{subsec:model}

\begin{figure}[t] 
    \centering
    \includegraphics[width= 0.95\columnwidth]{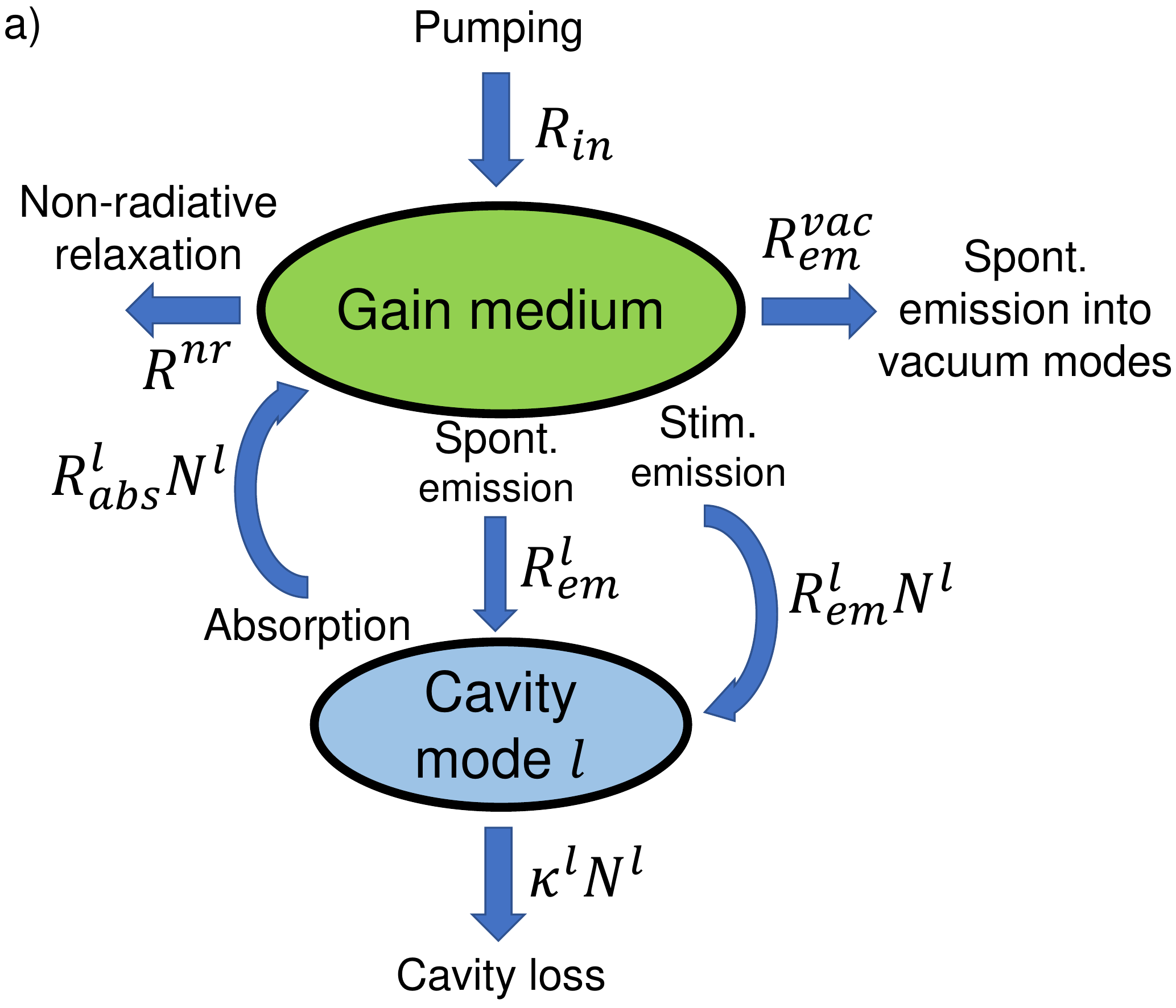} 
    
    \includegraphics[width= \columnwidth]{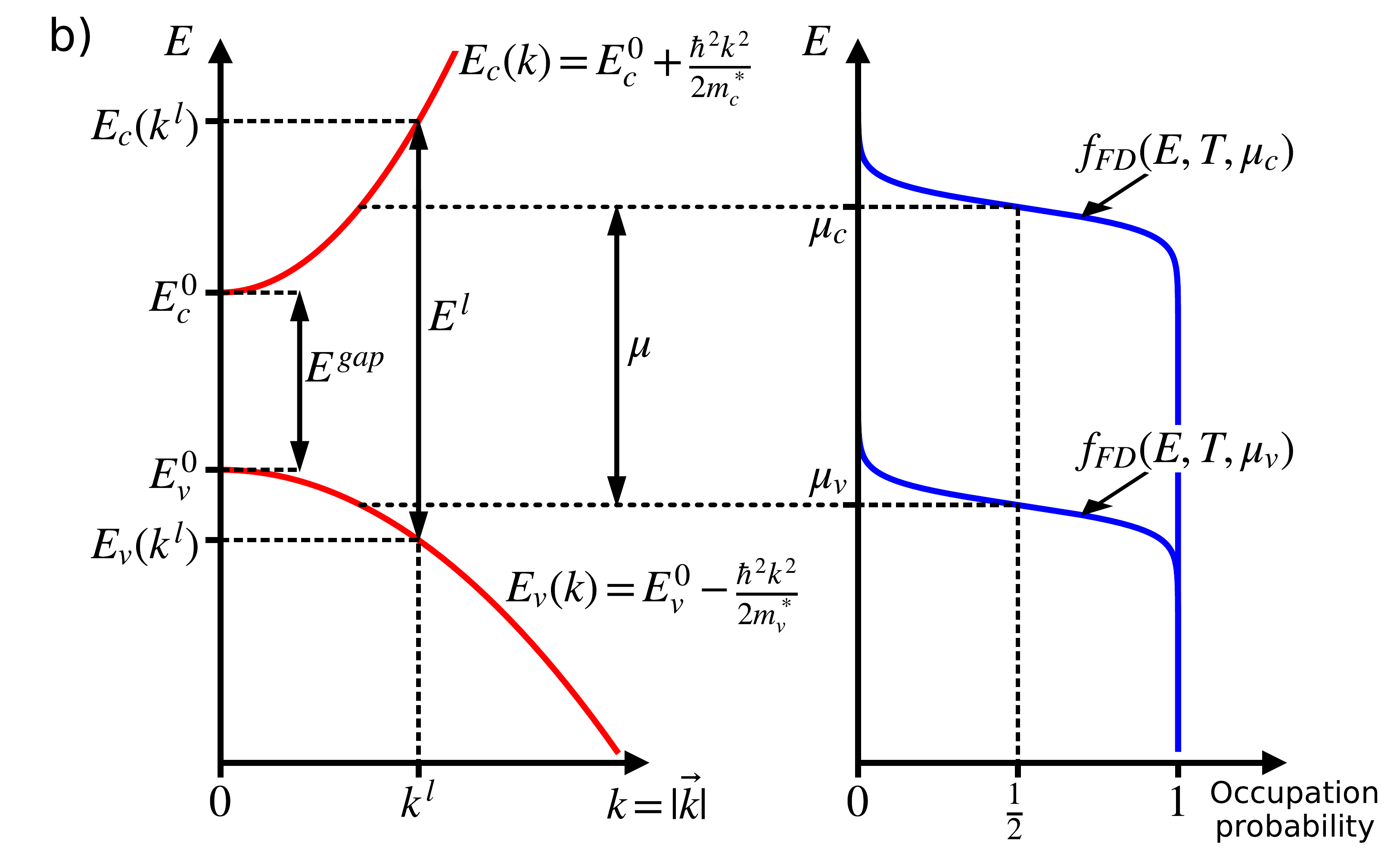}
    \caption{Scheme of the system and notations. Panel (a): flux of particles between the gain medium, the cavity, and the environment. Panel (b): semiconductor band structure as a function of the wavevector modulus (left) and distribution of the electrons in each band (right). {See the main text for a detailed description. }
}\label{fig:band_structure__distributions}
\end{figure}

Throughout this paper, we will focus on a piece of semiconductor placed in a cavity. We assume finite extension of the cavity so that photonics modes are {spectrally} discretized. We index them with $l=0,1,2...n_c$ corresponding to increasing energies. The various particle exchange pathways between the gain medium, the modes and the environment are shown on Fig. \ref{fig:band_structure__distributions} (a). In the cavity, photons in the $l$-th mode can be created or annihilated by the gain medium at the rates $R_{em}^l $ for spontaneous emission, $R_{em}^l N^l$ for stimulated emission and $R_{abs}^l N^l$ for absorption, where $N^l$ is the number of photons in the mode $l$. Alternatively, {radiative cavity losses occurs at the rate $\kappa^l N^l$}. In the semiconductor, excited electrons are created at the rate $R_{in}$ through pumping (indistinctly electrical or optical) 
{Conversely, relaxation can occur through the above-depicted emission in the cavity modes, through spontaneous emission into vacuum modes at the rate $R^{vac}_{em}$, or through non-radiative relaxation pathways (for example Auger effect) at the rate $R^{nr}$.}

In contrast with dye molecules, explicit forms of $R_{em}^l$ and $R_{abs}^l$ can be derived for semiconductors. Here we focus on an intrinsic direct bandgap semiconductor, indifferently 2 or 3-dimensional, and follow usual approximations \cite{coldren_corzine}. As sketched on Fig. \ref{fig:band_structure__distributions} (b), the conduction and heavy-hole valence band \footnote{Transitions between the valence light-hole and split-off bands and the conduction band are neglected.} are described by the isotropic dispersion $E_{c}(k)$ and $E_{v}(k)$ respectively, were $k$ stands for the wavevector modulus. Assuming that only vertical interband transitions are possible, a transition involving a photon in the mode $l$ with energy $E^l$ requires an electron and a hole with the same wavevector $\Vec{k}^l$ so that $E_c(k^l) - E_v(k^l) = E^l$. We also assume that the ground cavity mode energy is higher than the gap energy $E^0 > E^{gap}$. 

Interestingly, conduction electrons and valence holes close to the gap edges can be well described as free particles with an effective mass $m^*_{c/v}$, which leads to the simple parabolic band model $E_{c/v}(k) = E_{c/v}^0 \pm \frac{\hbar^2 k^2}{2m^*_{c/v}}$ with $E_{c/v}^0$ the energy minimum/maximum of the conduction/valence band and $\hbar$ the Planck constant. Analytical expressions for $E^l(k^l)$ can be derived, as well as for the density of states in each band $\rho_{c/v}(k)$ and the joint density of state $\rho_J(k)$ associated to the vertical transitions \cite{coldren_corzine}.

Next, we assume that the bands are in local thermodynamic equilibrium characterized by a Fermi-Dirac distribution with a common temperature $T$ and local chemical potentials $\mu_{c}$ and $\mu_{v}$, the so-called quasi-Fermi levels. In the case of electrical pumping, we have $\mu_{c} - \mu_{v}=eV$ where $V$ is the applied voltage. As the voltage increases, $\mu_{c}$ increases (resp. $\mu_{v}$ decreases) from the Fermi-level, so that their difference $\mu_{c} - \mu_{v}$ is controlled. It is also possible to define quasi-Fermi levels under optical pumping.

In this context, the spontaneous emission, stimulated emission and absorption rates for the mode $l$ can be written respectively as \cite{coldren_corzine}:

\begin{equation}\label{eq:emission_rate_detail}
\begin{split}
        R_{em}^l = g^l f_{FD}( E_c(k^l), T, \mu_c ) [ 1 - f_{FD}( E_v(k^l), T, \mu_v ) ]
\end{split}
\end{equation}

and 

\begin{equation}\label{eq:absorption_rate_detail}
    R_{abs}^l = g^l f_{FD}( E_v(k^l), T, \mu_v ) [ 1 - f_{FD}( E_c(k^l), T, \mu_c ) ],
\end{equation}

where $g^l$ is a {pumping-independent} transition rate and $f_{FD}(E,T,\mu) = 1/[ \exp{( \frac{E - \mu}{k_B T})} + 1 ] $ is the Fermi-Dirac distribution with $E$ the electron or hole energy, $k_B$ the Boltzmann constant and $\mu$ a quasi-Fermi levels. The microscopic expression of $g^l$ is given in Appendix \ref{app:fermi_golden_rule_g_i}. The right hand side of Eq. (\ref{eq:emission_rate_detail}) expresses that emission is proportional to the probability of finding an electron at the right energy in the conduction band and a corresponding hole in the valence band, and conversely for absorption in Eq. (\ref{eq:absorption_rate_detail}). 

Finally, we define the fraction of spontaneous emission into the mode $l$ through the generalized {$\beta$}-factor:

\begin{equation}\label{eq:beta_factor}
    \beta^l = \frac{ R_{em}^l }{ R_{em}^{vac} + \sum_j R_{em}^j  }.
\end{equation}

In a single cavity mode context, this dimensionless number characterizes the emission regime. A macroscopic laser corresponds to $\beta \to 0$ while a nanolaser corresponds to $\beta \to 1$. Indeed due to large (resp. small) mode volume,
a macroscopic (resp. nano-) laser is characterized by a low (resp. high) Purcell factor, so that spontaneous emission into the numerous vacuum modes (the mode $l$) is dominant. Reducing the volume further tends to reduce the cavity mode number. Still the cavity modes spacing and number can also be adjusted e.g. through engineering of the mirrors curvature for Fabry-Perot-like cavities. Hence the quantities $R_{em}^{vac}$ and $\sum_l R_{em}^l$ can be partially tuned independently. 

\subsection{Photon BEC in a lossless cavity with gain}\label{subsec:blackbody_approach}

{Bose-Einstein condensation is a property of an ensemble of bosons in thermodynamic equilibrium. Quantitatively, thermodynamic equilibrium means that a state at energy $E$ is occupied according to a Bose-Einstein distribution $1/[\exp( \frac{E - \mu}{k_B T}) -1 ]$ where $\mu$ is the chemical potential. Condensation may occur when the chemical potential approaches the ground state energy $\mu \to E^0$. }

{In the blackbody radiation, photons reach a thermodynamic equilibrium  due to walls acting as a reservoir. This equilibrium is characterized by a null photon chemical potential. Remarkably, Wurfel showed } that it is possible to introduce a photon chemical potential when dealing with stationary systems with gain \cite{wurfel1982chemical}. We reproduce here the reasoning for clarity. We start by assuming a perfectly lossless cavity, i.e. $\kappa^l=0$. In the steady-state regime, the balance between the spontaneous and stimulated emission processes and the absorption in the $l$-th photonic mode yields:

\begin{equation}\label{eq:balance_blackbody}
   R_{em}^l +  R_{em}^l N_{}^l = R_{abs}^l N_{}^l,
\end{equation}

where $N_{}^l$ is the number of photons in the $l$-th mode. It is readily seen that the photon number only depends on the ratio between the absorption rate and the emission rate $  R_{abs}^l  /  R_{em}^l  $. Given Eqs. (\ref{eq:emission_rate_detail}),(\ref{eq:absorption_rate_detail}) which assumes that the gain medium is in local thermodynamic equilibrium, simple algebra allows to recover the {Van Roosbroeck-Shockley relation \cite{van1954photon}: }

\begin{equation}\label{eq:neporent_etc}
    \frac{  R_{abs}^l  }{  R_{em}^l  } = \exp \bigg( \frac{E^l - \mu}{k_B T} \bigg),
\end{equation}

where $\mu = \mu_c - \mu_v$. From Eqs (\ref{eq:balance_blackbody}) and (\ref{eq:neporent_etc}), it follows that the photon number in the mode $l$ is given by: 

\begin{equation}\label{eq:photon_number_BE_blackbody}
    N_{}^l = \frac{1}{ \exp \big( \frac{E^l - \mu}{k_B T} \big) -1},
\end{equation}

namely a Bose-Einstein distribution with temperature $T$ and a chemical potential defined as the quasi-Fermi levels splitting. In the absence of pumping, the chemical potential is null and we recover the blackbody radiation distribution with the temperature of the semiconductor at equilibrium.

Finally, beyond the semiconductor model used here, we emphasize the key role of local thermodynamic equilibrium in each band under pumping to derive this result. Indeed, this appears as a sufficient condition on the gain medium to reach photons BEC. In particular, this explains why eq. (\ref{eq:neporent_etc}) can be written similarly for dyes molecules in terms of emission and absorption cross sections, a formula known as the Kennard-Stepanov relation \cite{kennard1918thermodynamics,kennard1926interaction,Ste57} {(sometimes also called the Neporent-McCumber relation, see Ref. \cite{band1988heller} and references therein)}. 
To summarize, the number of photons in a non-lossy cavity filled with a gain medium in local thermodynamic equilibrium can be described by a Bose-Einstein distribution with a non-zero chemical potential.

\subsection{The driven-dissipative regime of a lossy cavity with gain: Lasing or BEC ?}\label{section:lasing_or_BEC}

We now consider a cavity coupled to the environment through the loss rates $\kappa^l > 0$. Such a system composed of a gain medium and a cavity with radiative losses is usually considered to be a laser. A natural question then arises: what is the difference between Bose-Einstein condensation and lasing ? 

We repeat the analysis of the previous section using the same assumptions and notations, now accounting for cavity losses so that the system is in the driven-dissipative regime. The balance equation (\ref{eq:balance_blackbody}) becomes $R_{em}^l +  R_{em}^l N_{}^l = ( R_{abs}^l + \kappa^l ) N_{}^l$. The steady-state photon number in the mode $l$ can then be cast in the form \cite{coldren_corzine}:

\begin{equation}\label{eq:photon_number_mode_i_laser}
    N_{}^l = \frac{ R_{em}^l }{ \kappa^l - ( R_{em}^l - R_{abs}^l) }.
\end{equation}

In this last equation, the quantity $R_{em}^l - R_{abs}^l$ is better known as the net gain rate of the mode $l$. Hence, this simple model recovers that the mode $l$ starts to lase as the {net} gain compensates the {radiative} losses. So far, we have isolated a mode {and computed its occupation number} by expressing the balance between gain and losses. This approach is at first glance at odds with the study of the population of different modes in an equilibrium system. Nevertheless, we now cast this laser equation in a form that mimicks Eq.(\ref{eq:photon_number_BE_blackbody}). Upon factorization by $R_{em}^l$ and inserting the relation (\ref{eq:neporent_etc}) in eq. (\ref{eq:photon_number_mode_i_laser}), we find the alternative form \cite{torma2014PRA}:

\begin{equation}\label{eq:photon_number_BE_laser_approach}
    N_{}^l = \frac{1}{ \exp\left( \frac{E^l - \mu}{k_B T} \right) [1 + K_n^l(T,\mu)] -1 },
\end{equation}
 
where 

\begin{equation}\label{eq:knudsen_definition}
    { K_n^l(T,\mu) = \frac{\kappa^l }{R_{abs}^l(T,\mu) }  }
\end{equation}

is a dimensionless number often called Knudsen number in the context of transport phenomena and Boltzmann equation. The Knudsen number is given by the ratio of the absorption time  $\frac{1}{R_{abs}^l }$ by a characteristic time of the cavity, the residence time of a photon in the cavity $\frac{1}{\kappa^l }$. Hence, in the regime where a photon undergoes a large number of absorption and emission cycles during the residence time, the Knudsen number is small and the distribution (\ref{eq:photon_number_BE_laser_approach}) approaches the BE distribution of a non-lossy cavity. In other words, {the large number of
absorption and emission events enables the photons to thermalize with the semiconductor acting as a reservoir}. The Knudsen number appears to be the natural quantity that quantifies how thermalized is a mode. Importantly, note that a Knudsen number is associated to each mode, it is not a global quantity. We stress that some modes may be thermalized while others are not.

As a conclusion of this section, it is clear from eq. (\ref{eq:photon_number_BE_laser_approach}) that Bose-Einstein condensation of photons is a particular regime of lasing, in which (i) Eq. (\ref{eq:neporent_etc}) is satisfied for the gain medium and (ii) the Knudsen number is small for all the modes to ensure that they are all thermalized. In the remaining of this work, we will use "lasing" to refer indistinctly to Bose-Einstein condensation or standard out-of-equilibrium lasing. In addition, Eq. (\ref{eq:photon_number_BE_laser_approach}) provides an alternative point of view to interpret lasing. Indeed, while Eq. (\ref{eq:photon_number_mode_i_laser}) provides a good description of single mode lasing in a system with significant losses and gain, we anticipate that Eq. (\ref{eq:photon_number_BE_laser_approach}) will be more suited to the study of multimode phenomena in the thermalized regime.

\subsection{Knudsen number, thermalization degree, optical thickness, cooperativity { and photon number at transparency}  }

In the last section, we have introduced the Knudsen number $K_n^{l}$ of a mode $l$ as the absorption time divided by the residence time in the cavity. It takes small values in the thermalized regime. Its inverse, {that we note $D^{l}_{}$,} was called thermalization degree in Ref. \cite{nyman_nature_2018} or thermalization coefficient in Ref. \cite{hesten2018whenlessismore}. Its key role in photon Bose-Einstein condensation had been suggested \cite{klaers2010thermalization} and identified \cite{kirton_keeling_2016} in early papers. Here, we have shown how it appears naturally from laser rates equation in the context of an equilibrium distribution perturbed by the introduction of cavity losses. Let us now discuss alternative physical interpretations of the thermalization degree. We first note that it can be viewed as the effective cavity length $ L^{l}= c/\kappa^l $ divided by the absorption mean free path $l_{abs}^{l}= c /R_{abs}^l$. With this point of view, which is often used to discriminate between diffusive regime and ballistic regime in transport phenomena, we identify the degree of thermalization with the optical thickness $L^{l}/l_{abs}^{l}=D^{l}$. {Second, we} remind that the optical thickness is proportional to the cooperativity $C(N_a)$. This quantity had been initially introduced to characterize the absorption of a photon by an ensemble of $N_a$ atoms in a cavity in the context of non-linear optics in a cavity \cite{bonifacio1978optical}. It is currently used as a measure of the light-matter interaction in cavity quantum electrodynamics (CQED) \cite{marquier2017revisiting}. {Finally, the thermalization degree has been interpreted historically in laser physics as the photon number at transparency \cite{bjork1991yamamoto}. Here this follows from  Eq. (\ref{eq:photon_number_mode_i_laser}) when $R_{em}^l = R_{abs}^l$. Interestingly, this suggests to reinterpret some experiments featuring a high photon number at transparency as Bose-Einstein condensation of photons, see for example Ref. \cite{takemura2019lasing} for a semiconductor laser in a photonic crystal cavity.

\subsection{Lasing mode {in the BEC picture} }\label{sec:lasing_mode}

In the previous sections, we showed that the laser equation (\ref{eq:photon_number_BE_laser_approach}) giving the mode photon number has the structure of a Bose-Einstein distribution apart from a correction term given by $1+K_n^l(T,\mu)$. Hence, we can revisit the lasing transition in terms of Bose-Einstein distribution. 

We start with the laser point of view given by Eq. (\ref{eq:photon_number_mode_i_laser}). In this framework, {lasing in the mode $l$ occurs as the gain rate saturates when it approaches the loss rate $( R_{em}^l - R_{abs}^l) \to \kappa^l$. This is called gain clamping. In addition, finite losses require positive gain, that is, population inversion of the corresponding transition \footnote{Note that in contrast to a 2-level system, the sum of the valence and the conduction state occupation at a given wavevector is not necessarily unity, ie $f_{FD}( E_c(k), T, \mu_c ) + f_{FD}( E_v(k), T, \mu_v ) \in [0,2]$. This is due to effective mass imbalance between bands. Hence, transparency does not necessarily implies that $f_{FD}( E_c(k), T, \mu_c ) = f_{FD}( E_v(k), T, \mu_v ) =\frac{1}{2}$.}. }

We now place in the perspective of the generalized Bose-Einstein distribution. We start by writing (\ref{eq:photon_number_BE_laser_approach}) in a slightly different form \cite{torma2014PRA}:

\begin{equation}\label{eq:photon_number_BE_mu_effectif}
    N_{}^l = \frac{1}{ \exp\Big( \frac{E^l - \mu_{eff}^l(T,\mu)}{k_B T}\Big) -1 },
\end{equation}

where we have introduced an effective chemical potential $\mu_{eff}^l (T,\mu) = \mu - k_B T \log[1 + K_n^l(T,\mu) ]$. Here, we stress that this form enables to use the Bose-Einstein distribution which is an equilibrium concept in the nonequilibrium driven-dissipative regime. The effective chemical potential is composed of a term $\mu$ which accounts for the gain and a term $-k_BT \log[1+K_n]$ which accounts for the losses. The usual condition for Bose-Einstein condensation in the mode $l$ is then directly generalized as:

\begin{equation}
    \mu_{eff}^l(T,\mu) \to E^l. 
\end{equation}

{Here, the increase of the pump power is interpreted as increasing the quasi-Fermi levels splitting. Therefore $\mu$ converges toward a fixed value $\mu_{clp}$ defined as the solution of the implicit equation: }

\begin{equation}\label{eq:mu_clamping_implicit_equation}
    \mu_{clp} - k_B T \log[1 + K_n^l(T,\mu_{clp}) ] = E^l.
\end{equation}

{This saturation of $\mu$ corresponds to gain clamping in the BEC point of view.} In this last equation, the correction term is always negative. Hence, the quasi-Fermi levels splitting must exceed the transition energy $E^l$ to trigger lasing. This corresponds to population inversion. It highlights the importance to distinguish between the quasi-Fermi levels splitting and the effective chemical potential, since only the latter can be interpreted as the photon chemical potential.

We now focus on a multimode system. The usual laser textbook picture is the following \cite{coldren_corzine,milonni2010laser,svelto1998principles}: the gain curve is taken to be a bell-shaped function of frequency, while the frequency dependence of the mirrors losses is neglected. Lasing is thus expected to occur in the cavity mode with largest gain. This picture is at odds with the one of ideal Bose-Einstein condensation, which is expected to occur in the ground cavity mode.

We now revisit this issue using the Bose-Einstein picture given by Eq. (\ref{eq:mu_clamping_implicit_equation}). In the present multimode situation, each mode $l$ defines a different clamping value of the quasi-Fermi levels splitting, that we note $\mu_{clp}^l$. Single mode lasing takes place in the mode with the smallest $\mu_{clp}^l$. To gain further insight, we assume $K_n^l(\mu_{clp}^l) \approx K_n^l(E^l)$. The clamped quasi-Fermi levels splitting of each mode $l$ is simply given by:

\begin{equation}\label{eq:modechoice}
    \mu_{clp}^l \approx E^l + k_B T \log[ 1 + K_n^l(T,E^l) ].
\end{equation}

Interestingly, this expression is composed of two competing terms: on one hand, the mode energy favors lasing in low energy modes; on the other hand, it depends on the Knudsen number and favors lasing in highly thermalized modes. Therefore, without the second contribution coming from the cavity losses, we would recover the usual condensation on the ground mode. In practice, lasing in a mode above the ground mode is thus the signature of a system in which the modes have very different thermalization degrees. This discussion highlights that thermalization is primarily a modal property and not a system property. Indeed, as explained in Sec. \ref{subsec:blackbody_approach}, thermalization occurs between a mode and the reservoir, rather than between modes.

Finally, we note that some authors used lasing in the ground mode versus an excited mode as a criterion to distinguish between BE condensation and out-of-equilibrium lasing \cite{hesten2018whenlessismore,vlaho2019controlled}. While lasing in an excited mode is indeed a signature of nonequilibrium operation, Eq. (\ref{eq:modechoice}) shows that condensation in the ground mode is only the signature of a Knudsen number slowly varying from one mode to another, regardless of its absolute amplitude.

\subsection{Condensation versus lasing threshold}\label{subsec:Generalized_condensation_threshold}

In the previous section, we showed how to interpret the mode selected for lasing within a generalized Bose-Einstein condensation approach, stressing that condensation and lasing are two faces of the same coin. As a next step, it is natural to compare the definitions used for lasing threshold and for condensation threshold.

We first remind the lasing threshold definition. Many different criteria can be used to characterize lasing \cite{samuel2009recognize,carroll2021thermal}. Here, we consider the widely used condition based on an input/output curve. On Fig. \ref{fig:seuil_laser} (a), the number of photons $N^j$ in the cavity is plotted as a function of the injection rate of excited carriers, that we note $R_{in}$. On a linear scale,  $N^j$ turns suddenly from sublinear to linear on a small pumping range. The threshold is defined as the input rate of excited carriers $R_{in,LAS}$ when the linear slope is continued down to $0$ output rate (see {the dashed blue line on} Fig. \ref{fig:seuil_laser} (a)). This input rate is equal to the value of the losses, evaluated at gain clamping. Indeed, close to clamping, stimulated emission funnel all additional photons in the lasing mode. The losses are due to different mechanisms: the leakage through non-lasing cavity modes with rate $ \sum_{l\neq j} \kappa^l N^l(\mu_{clp}^j)$, the emission into vacuum modes ($R^{vac}_{em}(\mu_{clp}^j)$) and other non-radiative charge carrier relaxation processes ($R^{nr}(\mu_{clp}^j)$). The lasing threshold is thus given by:

\begin{equation}\label{eq:laser_threshold_rate}
    R_{in,LAS} = R^{nr}(\mu_{clp}^j) +R^{vac}_{em}(\mu_{clp}^j) + \sum_{l\neq j} \kappa^l N^l(\mu_{clp}^j).
\end{equation}

We now focus on the condensation threshold definition. In the literature on BEC in thermodynamic equilibrium, the BEC threshold is defined by the equality between the total number of particles and the number of particles in the excited states in the condensed phase \cite{pitaevskii2016bose}. First, note that in  photons BEC experiments, the number $N^l$ of photons in a mode $l$ cannot be measured directly. Still, the driven-dissipative regime enables to derive it from the measured flux $\kappa^l N^l$ and the knowledge of the loss rate $\kappa^l$. Second, note that in essence, this definition relies on the same idea as for a laser: beyond threshold, all additional photons will go to the condensed phase. As shown on Fig. \ref{fig:seuil_laser} (b), the condensation threshold is extracted graphically in a similar fashion as for the laser threshold when plotting the number of photons in the condensed mode versus the total number of photons in the cavity. The total number of photons in the cavity at threshold is then given by the sum of non-condensing modes population at clamping, namely $\sum_{l\neq j} N^l(\mu_{clp}^j)=N_{BEC}^{tot}$. Still, this procedure differs from the lasing threshold definition, as it is based on a number of photons in a cavity and not on a comparison of fluxes of input carriers and emitted photons. In particular, the nonradiative losses and the radiative losses into vacuum modes are not taken into account. Hence, the BEC definition leads to a smaller value of the threshold for the {quasi-Fermi levels splitting } than the lasing condition. The difference is not very large when the $\beta$ factor is close to 1 but may be very large when emission into vacuum modes dominates. This is illustrated in Figure \ref{fig:seuil_laser} (c) where it is seen that the thresholds can differ by orders of magnitude (in term of photons in the lasing mode).

To conclude, the choice of using the BEC or laser threshold has to be conducted carefully, as they can take very different values. To guide this choice, one should note that for the lasing threshold, both the input and emitted power must be monitored, while the emitted power spectrum is sufficient to determine the condensation one. As already encountered in the previous sections, this suggests that other than making a real difference between BEC and lasing, the "condensation" point of view is a framework suited to the study of the multimode character of the system, while the "lasing" one rather focus on its driven-dissipative aspect.

\begin{figure}[]
    \centering

    \includegraphics[width= \columnwidth]{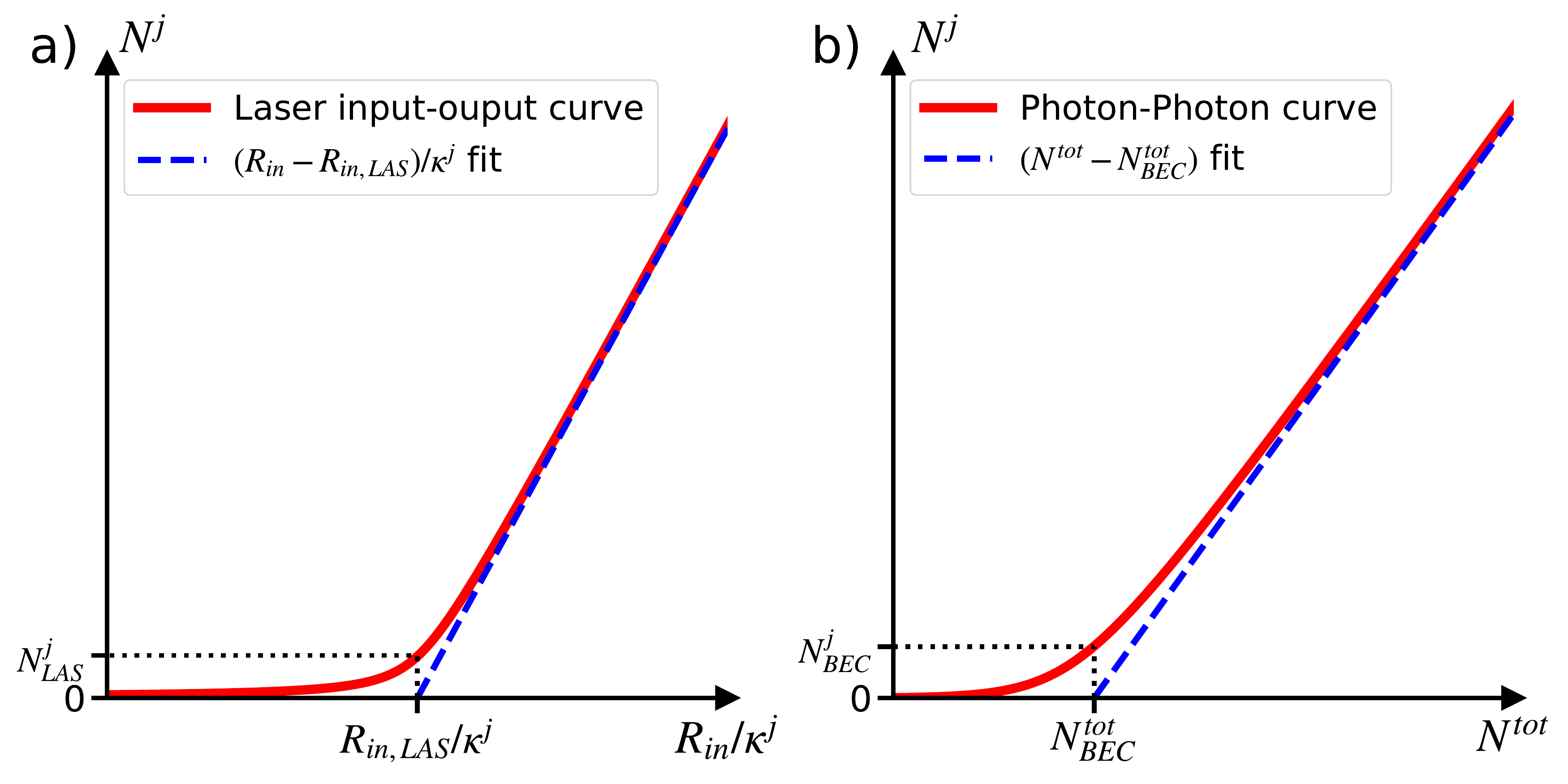}
        
    \includegraphics[width= \columnwidth]{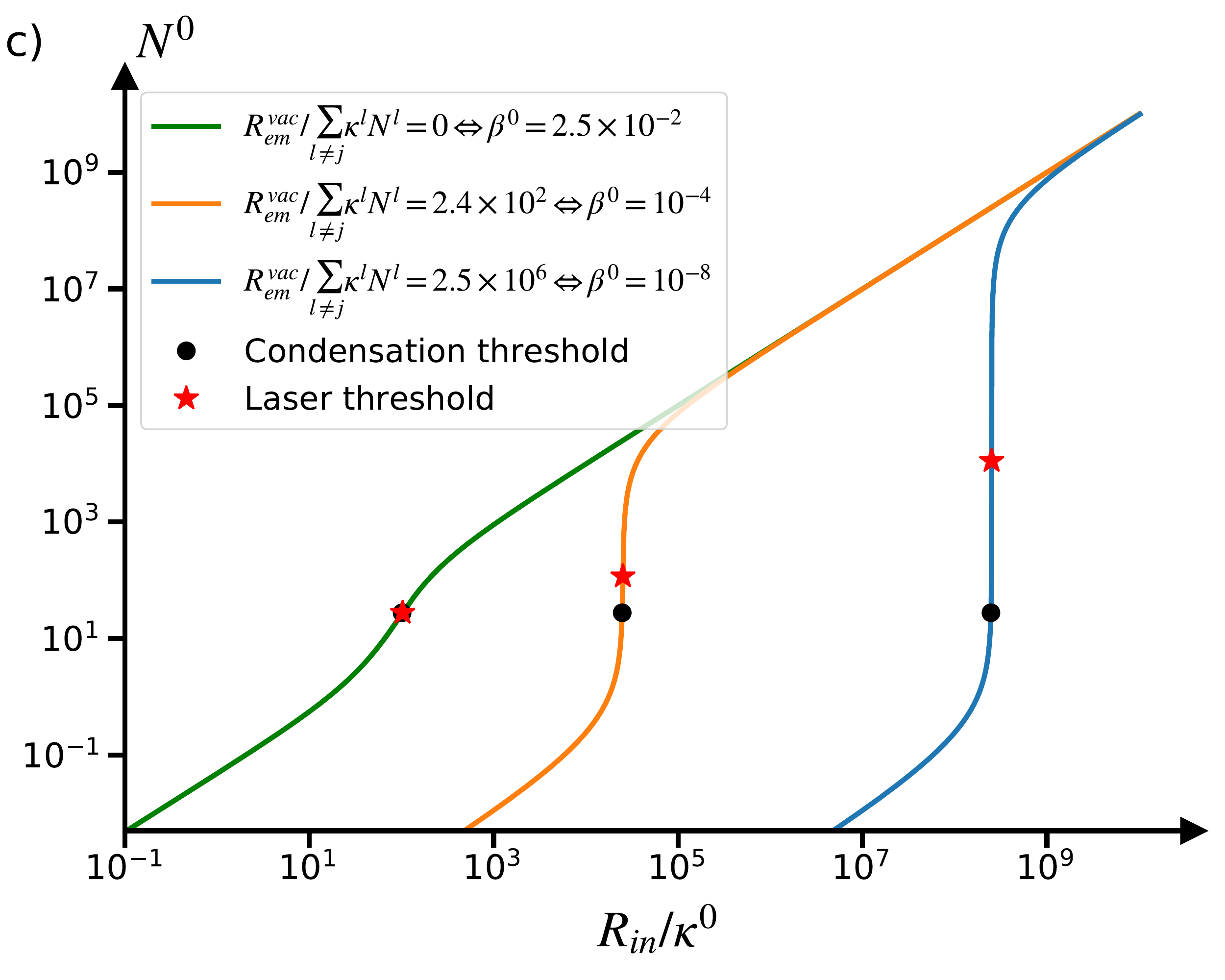}
     
    \caption{Panel (a): schematic input-output laser curve (red line) on a linear scale. The dashed blue line is a linear fit of the laser curve. Its intersection with the $N^j=0$ axis defines the laser pumping threshold $R_{in,LAS}$. The corresponding lasing mode photon number at threshold is noted $N^j_{LAS}$. Panel (b): schematic photon-photon curve of a multimode driven-dissipative BEC condensing in the mode $j$ (red line) on a linear scale. $N^{tot}= \sum_l N^l$ is the total number of photons in the cavity. The dashed blue line is a linear fit of the BEC curve. Its intersection with the $N^j=0$ axis defines the BEC threshold $N^{tot}_{BEC}$. The corresponding lasing mode photon number at threshold is noted $N^j_{BEC}$. Panel (c): comparison of the lasing mode photon number at laser and BEC thresholds, on input-output curves corresponding to different rates of spontaneous emission into vacuum modes. A constant cavity modes spacing is assumed so that their energy reads $E^l = E^0 ( 1 + 0.001\times l)$, with $E^0= 1.271$ eV. $\kappa^l$ and $g^l$ are assumed constant over the modes, with a ratio $g^l/\kappa^l = 10$. This enforces lasing in the ground mode. Non-radiative losses are neglected $R^{nr}=0$. To help considering the value of $R^{vac}_{em}(\mu_{clp}^j)/\sum_{l\neq j} \kappa^l N^l(\mu_{clp}^j)$, the corresponding value of $\beta^0$ is given in the legend. Other parameters are compiled in Appendix \ref{app:parameters}.
    }
    \label{fig:seuil_laser}
\end{figure}

\section{Experimental assessment of thermalization}\label{sec:spectrum}

In the previous section, we made a clear distinction between BEC and lasing using the thermalization degree of the modes. However, the thermalization degree cannot be measured directly. Indeed, it is proportionnal to the absorption rate $R^{l}_{abs}$, but only the net absorption rate $R^{l}_{abs}-R^{l}_{em}$ is given by a transmission measurement. In this section, we aim at finding observable quantities that depend {sharply} on the thermalization degree, enabling its assessment. We first analyze the emission spectrum under homogenenous pumping, which is the most common experimental practice, and find that this method may not be reliable. We then discuss spectral and spatial measurements under inhomogeneous pumping. We finally discuss the influence of band-filling on the thermalization degree.

\subsection{{Spectrum analysis}}\label{subsec:maxwell_boltzmann_spectrum}

In an ideally thermalized system, we saw in Section \ref{subsec:blackbody_approach} that the mode occupation follows a Bose-Einstein distribution $N_{}^l= 1/[\exp( \frac{E^l - \mu}{k_B T}) - 1]$. 
At low occupation numbers, the classical regime is recovered, namely, the BE distribution reduces to a Maxwell-Boltzmann distribution $N_{}^l\approx \exp( -\frac{E^l - \mu}{k_B T}) $. Hence, a common practice to prove thermalization consists in looking for a linear decay on a semilogarithmic plot of the spectrum \cite{bajoni2007,barland,kasprzak2006,weill2017thermalization}. Here, we compare this approach with the characterization based on the Knudsen number.

In the classical regime, the generalized BE distribution eq. (\ref{eq:photon_number_BE_laser_approach}) becomes:

\begin{equation}\label{eq:maxwell_boltzmann_thermalization_degree}
    N_{}^l \approx  \frac{ \exp( -\frac{E^l - \mu}{k_B T}) }{ 1 + K_n^l} .
\end{equation}

It is readily seen that an exponential decay of the cavity photons spectrum is observed in two cases: (i) the Knudsen number of all the modes is much lower than $1$, and (ii) the Knudsen number is constant over the modes, whatever its value. In the second case, despite an exponential behaviour of the spectrum, the Knudsen number may take values $\gtrsim 1$ indicating a non thermalized system. 

Beyond the classical regime, it is noteworthy that this issue persists in the quantum degenerate regime. Indeed, according to Eq. (\ref{eq:photon_number_BE_mu_effectif}), the generalized BE distribution with constant Knudsen number $K_n$ simplifies in an equilibrium BE distribution with the effective chemical potential $\mu_{eff} = \mu - k_B T \log[1 + K_n]$ \cite{torma2014PRA}. All in all, it means that spectrum analysis with homogeneous pumping in order to quantify the thermalization may not be reliable. In particular, we note in Appendix \ref{app:D_l_independent_l} that devices featuring a large, planar and homogeneously pumped cavity are likely to feature a nearly constant Knudsen number. This may explain the BE-like spectra observed in optically \cite{bajoni2007,lagoudakis2012crossover} and electrically \cite{barland} pumped large area VCSELs.

\subsection{Inhomogeneous pumping}\label{subsec:inhomogeneous_pumping} 

An interesting signature of thermalization can be observed when using an inhomogeneous pumping with a beam {or injection area} much smaller than the cavity. Indeed, the pumped part of the gain medium emits photons isotropically through spontaneous emission. These photons can be reabsorbed efficiently everywhere in a thermalized system. As a consequence, the gain is homogeneous in the cavity despite a localized pumping.  
To describe this effect, it is necessary to include additional rate equations describing locally the gain medium population \cite{kirton_keeling_2016,hesten2018whenlessismore,walker2019collective}. While this goes far beyond the scope of the present work, we give a hint of the complexity of this case by writing how the photon occupation number is modified. The balance equation (\ref{eq:balance_blackbody}) with the losses $\kappa^l$ for a mode $l$ has to be integrated over the gain medium volume (also called active  volume) $V_{act}$, namely $ \int_{V_{act}} d^3\Vec{r}\, \big[ R_{em}^l(\Vec{r}\,) +  R_{em}^l(\Vec{r}\,) N_{}^l \big] = N_{}^l \int_{V_{act}} d^3\Vec{r}\, \big[ R_{abs}^l(\Vec{r}\,) + \kappa^l/V_{act} \big]$ where the rates are now defined locally. In particular, the local Knudsen number is $K_{n}^l(\Vec{r}) = \kappa^l/ [R_{abs}^l(\Vec{r}\,) V_{act}]$. The photon number in the mode $l$ then becomes:

\begin{equation}\label{eq:BE_generalized_spatial}
    N^l = \frac{1}{  \frac{ \exp( \frac{E^l}{k_B T} ) }{ \big\langle \exp( \frac{\mu(\Vec{r}\,)}{k_B T} ) \big\rangle_{}^l }  \Big[ 1 + \big\langle  K_{n}^l(\Vec{r}\,)  \big\rangle_{}^l \Big] -1 },
\end{equation}

where $\langle A(\Vec{r}\,) \rangle_{}^l =  \int_{V_{act}} d^3\Vec{r} \, R_{abs}^l(\Vec{r}\,) A(\Vec{r}\,) / \int_{V_{act}} d^3\Vec{r} \, R_{abs}^l(\Vec{r}\,) $ is a spatial average weighted and normalized by the absorption rate. While the global distribution still appears as a generalized BE distribution, additional complexity is brought by the spatial average. {In particular, the weighting by the local absorption now gives a modal dependence to the quasi-Fermi levels splitting. An enhanced sensitivity to imperfect thermalization is thus expected. Experimentally, it was reported in Ref. \cite{klaers2010thermalization} a departure from the ideal BE distribution of high energy modes occupation, while the thermalization degree was tuned down. Given the small extension of the optical pump used compared to the large extension of these high energy modes, this is in good qualitative agreement with our considerations. A similar observation has also been made in Ref. \cite{torma2018nature} for plasmon-polaritons. }

{Beside spectrum analysis, we eventually mention two other types of measurements that reveal efficiently the thermalization of the system with inhomogeneous pumping. The first consists in measuring the size of the condensate as a function of the size of the pumping beam. When the system is well thermalized, the condensate size is invariant, while it follows the size of the spot in the opposite case. This type of measurement has been reported \cite{marelic2015experimental}. In the same fashion, the spatial position of the condensate in a trap can be compared with the position of the pump beam. As the pump is moved away from the center of the trap, the longer condensation keeps occurring in the center, the higher the thermalization rate. This measurement has been reported in Ref. \cite{klaers2010thermalization,schmitt2015dynamics}.}

\subsection{Thermalization and saturation at high pumping}\label{sec:saturatio_high_pumping}

In the last subsection, we discussed how the thermalization of a system can be probed with inhomogeneous pumping. {Noteworthy, this has been done as if the thermalization of a mode was a general quantity, independent on the pumping strength. 
Here, we discuss how the thermalization evolves as the system is driven} toward the degenerate regime through strong pumping. The key issue is simple: thermalization is ensured by absorption and reemission; if the gain medium is highly pumped and approaches saturation, absorption is reduced and hence thermalization decreases.  

\begin{figure}[htbp]
    \centering
    \includegraphics[width= 0.5\textwidth]{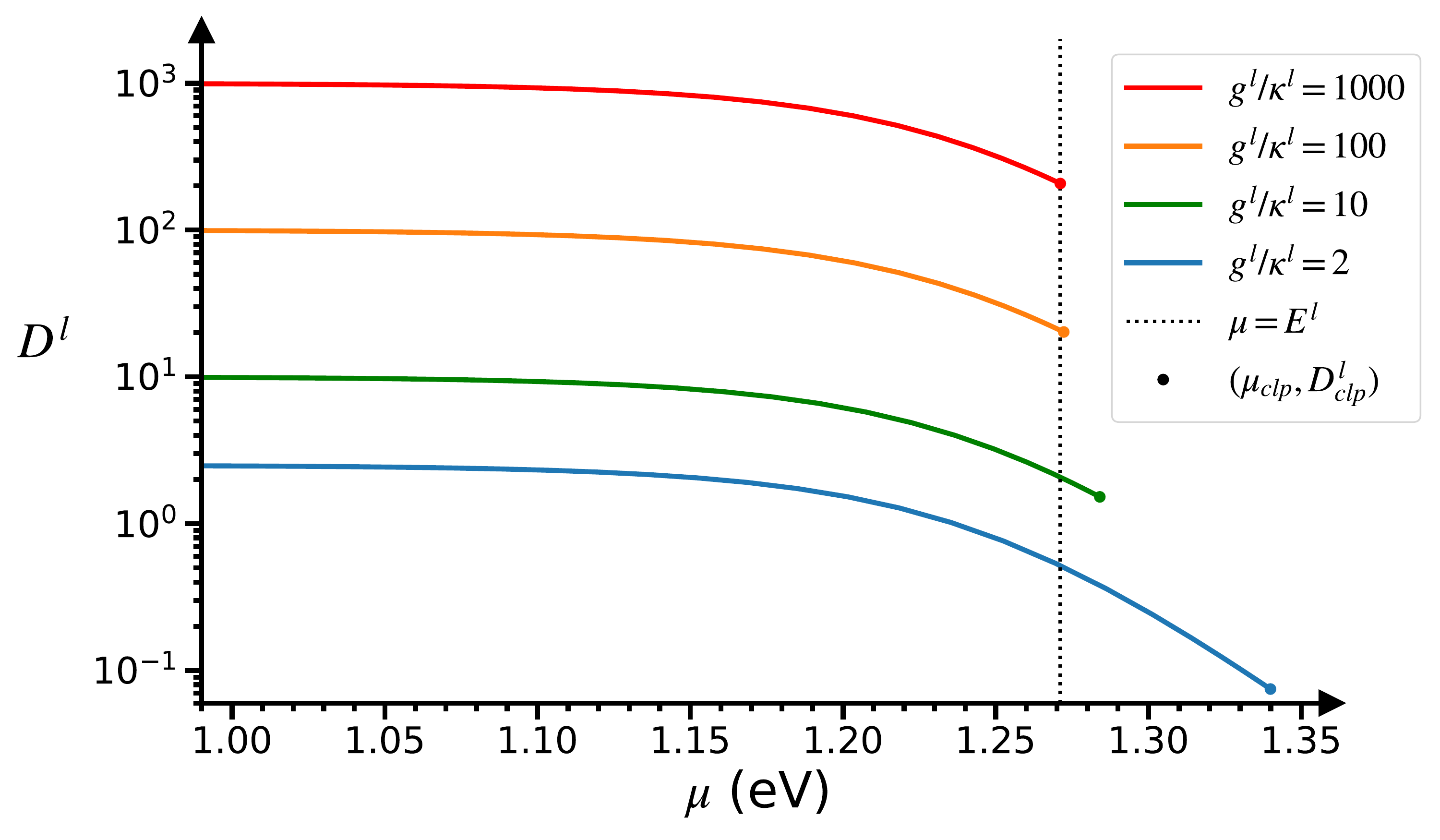}
    \caption{ Variation of the thermalization degree $D_{}^l$ of a mode $l$ as a function of the quasi-Fermi levels splitting $\mu$, for various ratio $g^l/\kappa^l$. The colored dots indicate clamping as defined in Eq. (\ref{eq:mu_clamping_implicit_equation}), at the quasi-Fermi levels splitting $\mu_{clp}$ and the thermalization degree $D^l_{clp} = D_{}^l(\mu_{clp})$. The vertical dark dashed line indicates transparency, namely $\mu = E^l$. The energy of the mode is $E^l= 1.271$ eV. Other parameters are compiled in Appendix \ref{app:parameters}.
    }
    \label{fig:Dth_mu}
\end{figure}

Inserting Eq. (\ref{eq:absorption_rate_detail}) into Eq. (\ref{eq:knudsen_definition}), the dependence of the thermalization degree on the quasi-Fermi levels (that is pumping) reads: 

\begin{equation}\label{eq:thermalization_degree_split}
     D_{}^l= \frac{g^l}{\kappa^l}  f_{FD}( E_v(k^l), T, \mu_v ) [ 1 - f_{FD}( E_c(k^l), T, \mu_c ) ].
\end{equation}

On Fig. \ref{fig:Dth_mu}, we show the evolution of the thermalization degree of a mode $l$ as a function of the quasi-Fermi levels splitting $\mu$ \footnote{As in an intrinsic semiconductor $\mu_c$ and $\mu_v$ are uniquely related, so that $\mu$ can directly be used as control parameter.} for various $g^l / \kappa^l$. At low pumping, 
filling of the conduction band (and accordingly depletion of the valence band) is negligible so that $D_{}^l = g^l / \kappa^l$. When increasing the quasi-Fermi levels splitting, the thermalization degree decreases significantly. At clamping (colored dot), the fall is about a  multiplication factor $1/5$ at high $g^l / \kappa^l$, and more than $1/10$ at low $g^l / \kappa^l$. In the first case, corresponding to a well thermalized mode, clamping occurs right over transparency (dark dashed vertical line). In a two-level system, transparency corresponds to an occupation probability of $1/2$ of the upper and lower level, so that the product of the levels occupation is $1/4$. Here, the slightly different value is due to the asymmetry of the bands of our semiconductor model (see Appendix \ref{app:parameters}). In the low mode thermalization case, a large inversion population is needed for lasing, that occurs well above transparency. The occupation probability of the conduction band is then much greater than $1/2$, and conversely for the valence band. Hence, the degree of thermalization is significantly decreased compared to the near-equilibrium case.

In summary, reliable assessments of the system thermalization should be made in the degenerate regime due to this dependence of the thermalization degree dependence on pumping.

\section{Intensity fluctuations: are they a BEC signature ?}\label{sec:intensity_fluctuations}

In the previous part, we showed that the spectrum is a quantity that can reveal the thermalization of the system, but which needs to be analyzed and probed with care. In this section, we investigate the intensity fluctuations as an alternative observable to distinguish between the BEC and the out-of-equilibrium laser regimes.

\subsection{Context}

{In the textbook picture of out-of-equilibrium lasing, coherence sets up right at the lasing threshold \cite{coldren_corzine}. {Above threshold, the intensity fluctuations are ruled by Poissonian statistics resulting in a second order correlation function at zero-time  delay $g^{(2)}(0)=1$.} On the contrary, earlier works on intensity fluctuations in the BEC regime predicted \cite{klaers2012PRL} and then measured \cite{schmitt2014PRL} super-Poissonian statistics for light well-above condensation threshold. This thermal regime, characterized by $g^{(2)}(0)=2$, was found to extend deeply in the condensed phase, before the crossover to the usual Poissonian light was recovered}. This ask the question whether large fluctuations are a signature of BEC.   

While the picture described in the last paragraph suggests studying the fluctuations according to the thermalization degree, other parameter have to be taken into account. In Refs. \cite{klaers2012PRL,schmitt2014PRL}, it has been pointed out that the reservoir size has an important influence on fluctuations. For large reservoirs, the gain medium can be loosely thought as an infinite reservoir{, recovering grand-canonical ensemble conditions. The large condensed mode photon number fluctuations, comparable to its mean value even above condensation threshold, are then identified to the so-called grand-canonical fluctuation catastrophe \cite{kocharovsky2006fluctuations}. On the contrary, fluctuations become limited when the reservoir excitations number is smaller than the mean photon number. 

Besides the role of the volume, it has been shown that the $\beta$-factor has a strong influence on the fluctuations for  micro- and nano-lasers \cite{vandruten2000,hess2000thermal}. While macroscopic lasers with low $\beta-$factor show the usual steep crossover between thermal and Poissonian light at lasing threshold, in high-$\beta$ devices the crossover is slow and occurs well above threshold. Knowing that the perfect equilibrium approach of Ref. \cite{klaers2012PRL} assumed a $\beta = 1$ cavity, this rather suggests that intensity fluctuations could be independent on the system thermalization, at least in the nanolaser limit.

In summary, assessing the role of thermalization on fluctuations requires to carefully control both the effect of the size of the reservoir and the $\beta$-factor. With that many degrees of freedom, it is a theoretical challenging task. Methods to calculate the photon number distribution like master equations for the lasing mode photon number \cite{rice1994carmichael,scully1999quantum,klaers2012PRL,kirton_killing_2015} or stochastic rate equations \cite{mork_lippi_2018,mork2020efficient,verstraelen2019temporal,walker2020natCom} can hardly been solved numerically. In the next subsections, we proceed to a simpler investigation, focusing only on the second-order coherence at zero-time delay $g^{(2)}(0)$. We calculate this quantity by studying the small photon number deviations over the steady state in the Langevin approach. Interestingly, we note that for this quantity, this approach showed good agreement with a more rigorous stochastic rate equations model \cite{mork_lippi_2018}. Thus, this allows to get accurate and analytical insights for an observable easily accessible experimentally.

\subsection{Second-order coherence at zero delay time }\label{sec:modele_langevin}

We base our investigation on the dynamical evolution equation of the photon number $N_{}^j(t)$ in the cavity mode $j$ that is lasing. For simplicity, we neglect the influence of other cavity modes on the system dynamics{, and of non-radiative losses}. Hereafter, we omit the mode superscript $N_{}^j = N_{}$. Following the notations of Fig. \ref{fig:band_structure__distributions} (a) for the various exchange pathways between the cavity, the gain medium and the environment, the dynamical evolution equation for $N_{}(t)$ is given by:

\begin{equation}\label{eq:dyna_N0_monoo}
    \frac{d N_{}}{dt} = - \kappa N_{} + [R_{em}(N_e) - R_{abs}(N_e) ] N_{} +  R_{em}(N_e),
\end{equation}

where $N_e(t)$ is the number of excited electrons in the gain medium. We switched from the variables $\mu_c$ and $\mu_v$ to $N_e$ for simplicity. 
In semiconductors gain media, the excited electrons dynamics is usually commensurable with the one of the cavity photons \cite{coldren_corzine}. Hence the corresponding evolution equation of $N_e(t)$ must be taken into account. According to Fig. \ref{fig:band_structure__distributions}, it yields:

\begin{equation}\label{eq:dyna_Ne_monoo}
    \frac{d N_e}{dt} = R_{in} - [R_{em}(N_e) - R_{abs}(N_e) ] N_{}  - \frac{R_{em}(N_e) }{\beta},
\end{equation}

where we used that $R^{vac}_{em}= (1/\beta - 1) R_{em}$ as follows from Eq. (\ref{eq:beta_factor}). {In the following, $\beta$ will be assumed to be independent on pumping, as usual in laser physics \cite{coldren_corzine}.} All in all, these 2-coupled rate equations correspond to a standard class-B model broadly used to describe the dynamics of most semiconductor single mode lasers \cite{coldren_corzine}. 

We now note the steady-state solutions of equations (\ref{eq:dyna_N0_monoo}),(\ref{eq:dyna_Ne_monoo}) as $N_{ss},\, N_{e,ss}$, respectively. We also introduce the small deviations $\delta N_{}(t),\, \delta N_e(t)$, with $|\delta N_{}| \ll N_{ss}$ and $|\delta N_{e}| \ll N_{e,ss}$.
We then linearize Eqs. (\ref{eq:dyna_N0_monoo}),(\ref{eq:dyna_Ne_monoo}) to first order in these parameters. The noise due to the quantization of the emission, absorption, pumping and loss process is finally added to each equation through the respective stochastic terms $F_{p},\, F_{e}$. We obtain the following coupled Langevin equations:

\begin{equation}\label{eq:dyna_delta_N0_monoo}
    \frac{d \delta N_{}}{dt} = - \gamma_{pp} \delta N_{} + \gamma_{pe} \delta N_e + F_{p}
\end{equation}

and

\begin{equation}\label{eq:dyna_delta_Ne_monoo}
    \frac{d \delta N_e}{dt} = -\gamma_{ep} \delta N_{} - \gamma_{ee} \delta N_e + F_e,
\end{equation}

where we have defined the short-hands $\gamma_{pp}= -[R_{em}(N_{e,ss}) - R_{abs}(N_{e,ss}) ] + \kappa$, $\gamma_{pe} = N_{ss} \partial_{N_e} [R_{em} - R_{abs} ] (N_{e,ss}) +  \partial_{N_e} R_{em} (N_{e,ss})$, $\gamma_{ep}= [R_{em}(N_{e,ss}) - R_{abs}(N_{e,ss}) ]$, $\gamma_{ee} = N_{ss} \partial_{N_e} [R_{em} - R_{abs} ] (N_{e,ss}) +  (1/\beta) \partial_{N_e} R_{em} (N_{e,ss})$. In addition, the stochastic terms verify the usual correlations properties $\langle F_x(t_1) F_y(t_2) \rangle = 2S_{xy} \delta(t_1 - t_2)$ with $x,y \in (p,e)$ \cite{coldren_corzine}, where the expressions of the $S_{xy}$ are given in Appendix \ref{app:g_2_complet}. 

In this linearized Langevin approach, the second-order intensity correlation $g^{(2)}(0) =  \langle N_{}(0) [N_{}(0) - 1] \rangle / N_{ss}^2 $ is given by:

\begin{equation}
    g^{(2)}(0) = 1 -  \frac{1}{ N_{ss} }  + \frac{\langle \delta N_{}(0)^2 \rangle   }{  N_{ss}^2 }.
\end{equation}

A detailed expression is then obtained by Fourier transforming (\ref{eq:dyna_delta_N0_monoo}),(\ref{eq:dyna_delta_Ne_monoo}), so that the problem can be reformulated into a matricial form which is easy to invert. The full result, well-known in the literature \cite{coldren_corzine, mork_lippi_2018,fedyanin}, is given in Appendix \ref{app:g_2_complet}. Simple asymptotic expressions can be written for limiting values of some parameters, as discussed in the next subsection. 

\subsection{Results}

We first focus on the usual macroscopic laser limit $\beta \to 0$. From the full expression in Appendix \ref{app:g_2_complet}, simple algebra show that the second-order coherence at zero delay time  reduces to:

\begin{equation}\label{eq:g2_beta_to_0}
    g^{(2)} (0) = 1 +  \frac{1}{ 1 + \big( \frac{ N_{ss} }{ N_{LAS}^{} } \big)^2 },
\end{equation}

where $N_{LAS}^{}$ is the photon number at lasing threshold. The coherence threshold is defined at $g^{(2)} (0)=1.5$ corresponding to $N_{} = N_{LAS}^{}$. Going straight to the point, this equality between the coherence and laser thresholds does not allow for a distinction between standard laser and photons BEC in this limit. Indeed here the crossover from thermal to Poissonian statistics always occurs at lasing threshold, \textit{regardless of the thermalization degree}.

We now focus on the opposite, "nanolaser" limit  $\beta \to 1$, where most of the spontaneous emission goes into the single cavity mode. The second-order coherence at zero delay time  now follows the asymptotic behaviour \cite{klaers2012PRL,verstraelen2019temporal,vandruten2000,hess2000thermal,fedyanin}:

\begin{equation}\label{eq:g2_fedyanin}
    g^{(2)} (0) = 1 +  \frac{1}{ 1 + \big( \frac{ N_{ss} }{ N_{CO} } \big)^2 },
\end{equation}

where the coherence threshold is now given by $N_{CO} \approx \Big[ \frac{R_{em}(N_{ss}= \infty)}{\partial_{N_e} [R_{em} - R_{abs}](N_{ss}= \infty)} \Big]^{1/2}$. \textit{It is seen that $N_{CO}$ presents no explicit dependence on the thermalization degree}. Coming back to our initial question, we conclude that the study of intensity fluctuations through the second-order coherence at zero delay time  does not provide a mean to distinguish between the out-of-equilibrium laser and the BE condensation regimes.

Finally, we discuss the consequence of this conclusion on the grand canonical fluctuation catastrophe. In the nanolaser limit, the coherence threshold is not given by the laser nor the BEC threshold. In particular, for realistic parameter values, the coherence threshold is shifted to much stronger pumping values than the laser threshold \cite{vandruten2000,hess2000thermal,fedyanin} and the BEC threshold \cite{klaers2012PRL}. Hence, there is a lasing/BEC regime with large fluctuations between these two thresholds. It is possible to attribute this regime to grand canonical fluctuations. Indeed, it has been shown \cite{schmitt2014PRL} that the coherence threshold square $N_{CO}^2$ corresponds to an effective number of excited carriers in the gain medium. Therefore, in the range between the condensation and the coherence thresholds, the gain medium is large compared to the photon gas and can be considered to be an infinite reservoir. Remarkably, we find that the concept of grand canonical fluctuations is not restricted to equilibrium BE condensation but can be extended to non-equilibrium systems.

\section{Conclusion}

To summarize, we have explored the photon Bose-Einstein condensate regime for semiconductors in a cavity. Owing to the explicit form of the gain for semiconductors and the extensive body of knowledge for semiconductors lasers, this system is a very convenient playground which provides a theoretical framework to discuss both lasing and condensation. Using the Van Roosbroek-Schockley relation, we have shown that the photon Bose-Einstein condensation in the driven dissipative regime is a particular case of the lasing regime. The theoretical framework also enables to compare the definitions of threshold used either for condensation or for lasing.  A  Knudsen number emerges naturally from the analysis to characterize thermalization.  We have discussed its close connection with other quantities introduced in different contexts such as thermalisation degree, optical thickness and cooperativity.  Equipped with this theoretical figure of merit to quantify thermalization, we have analysed different experimental  procedures to assess thermalization and put forward their strengths and limitations. Finally, we have explored the connection between the intensity fluctuations and the emission regime. Large fluctuations are a priori expected to be a signature of the grand canonical regime typical of the equilibrium condensation. However, using a Langevin analytical model of the fluctuations in the driven-dissipative regime, we showed that the coherence threshold does not depend on the thermalization degree, both for large and small $\beta$-factors. 

In this paper, we have explored the stationary regime of a single BEC. The semiconductor platform appears to be a very fruitful playground to study BEC physics. An interesting direction for future work is to revisit in the BEC regime recent results obtained with semiconductor cavities such as topological lasers \cite{bahari2017nonreciprocal,st2017lasing,parto2018edge}, chiral emission \cite{carlon2019optically}, nonlinearities \cite{hamel2015spontaneous} including superfluidity \cite{keijsers2020steady}. The platform is also well suited to further explore the dynamical behavior of BEC \cite{gladilin2020classical}. Also, the analysis of the fluctuations has revealed an interesting regime for micro and nanolasers above the lasing threshold and below the coherence threshold which can be viewed as grand-canonical fluctuations in non-equilibrium systems. This calls for more detailed studies of this phenomenon in the framework of open systems. It may provide new experimental platforms for the study of nonequilibrium statistical phenomena.

\begin{acknowledgments}

We are grateful to Gian Luca Lippi for helpful discussions. This work is supported by the French National Agency (ANR) (ANR-17-CE24-0046). J.-J.G. acknowledges the support of Institut Universitaire de France (IUF).

\end{acknowledgments}

\appendix

\section{Microscopic expression of $g^l$}\label{app:fermi_golden_rule_g_i}

According to the Fermi golden rule, it is shown that $g^l$ can be cast into the form \cite{coldren_corzine}:

\begin{equation}\label{eq:regle_d_or}
    g^{l} = \frac{2\pi}{\hbar} [\hbar \Omega]^2 \rho_J V_{act} \Gamma^{l},
\end{equation}

where $\hbar \Omega$ is the projected light-matter coupling Hamiltonian between a single vertical transition and a plane wave, $V_{act}$ is the volume of the gain medium (also called active medium) and $\Gamma^{l}$ is the overlap integral between the gain medium and the cavity mode. 
The spatial structure of the mode electric field is thus fully contained in $\Gamma^{l}$ which is thus mode dependent.

\section{Thermalization degree in devices featuring a large, planar and homogeneously pumped cavity}\label{app:D_l_independent_l}

Plugging Eq. (\ref{eq:absorption_rate_detail}) into Eq. (\ref{eq:knudsen_definition}), the thermalization degree has the form: 

\begin{equation}\label{eq:thermalization_degree_split_app}
     D_{}^l= \frac{g^l}{\kappa^l}  f_{FD}( E_v(k^l), T, \mu_v ) [ 1 - f_{FD}( E_c(k^l), T, \mu_c ) ],
\end{equation}

that is a pump-independent term $g^{l}/\kappa^l$ and a pump dependent term corresponding to the product of the Fermi-Dirac distributions.

According to the Appendix \ref{app:fermi_golden_rule_g_i}, the pump-independent term reads $g^{l}/\kappa^l = \frac{2\pi}{\hbar} [\hbar \Omega]^2 \rho_J V_{act} \Gamma^{l} /\kappa^l$. In semiconductor devices featuring a large and planar cavity, the measurable spectrum typically extends over $\sim 40$ meV  due to detection angle limitation and high refractive index material \cite{barland}. The variations of the transition matrix element $\Omega^l$, the joint density of states $\rho_J$ and the mirror loss rate $\kappa^l$ are negligible in this range \cite{coldren_corzine}. Also, as the in-plane part of the modes is nearly a plane wave, the overlap integral $\Gamma^l$ is constant. Hence, the pump-independent part of the thermalization degree $g^{l}/\kappa^l$ should thus be constant over the modes to a good approximation.

We now focus on the pump dependent term, which describes the saturation of the absorption through pumping (see Section \ref{sec:saturatio_high_pumping}). Rigorously speaking, this term has always some dependence on the modes, since saturation of the absorption is greater for low energy modes. Still, it is showed on Fig. \ref{fig:BE_vs_gen_BE} that the impact of this dependence is almost unnoticeable on the emitted spectrum for values of $g/\kappa$ as low as $\sim 4$. Noting that lasing is prevented if $g/\kappa < 1$ \footnote{This follows from the gain clamping condition $[ R_{em}^l(\mu_{clp}) - R_{abs}^l(\mu_{clp})] = \kappa^l$. The left hand side of this expression is bounded by the full inversion value of the gain $[R_{em}^l(\mu_{clp}\to \infty) - R_{abs}^l(\mu_{clp}\to \infty)] = g^l$. Losses that exceeds this bounds prevents gain clamping to occur, and thus lasing. Hence, the condition for lasing reformulates in $g^l /\kappa^l > 1$.}, the range over which the absorption saturation term has significant influence is quite narrow. As a conclusion, devices featuring a large, planar and homogeneously pumped cavity are expected to be well described by a constant Knudsen number over the modes.

\begin{figure}[thbp]
    \centering
    \includegraphics[width= 0.5\textwidth]{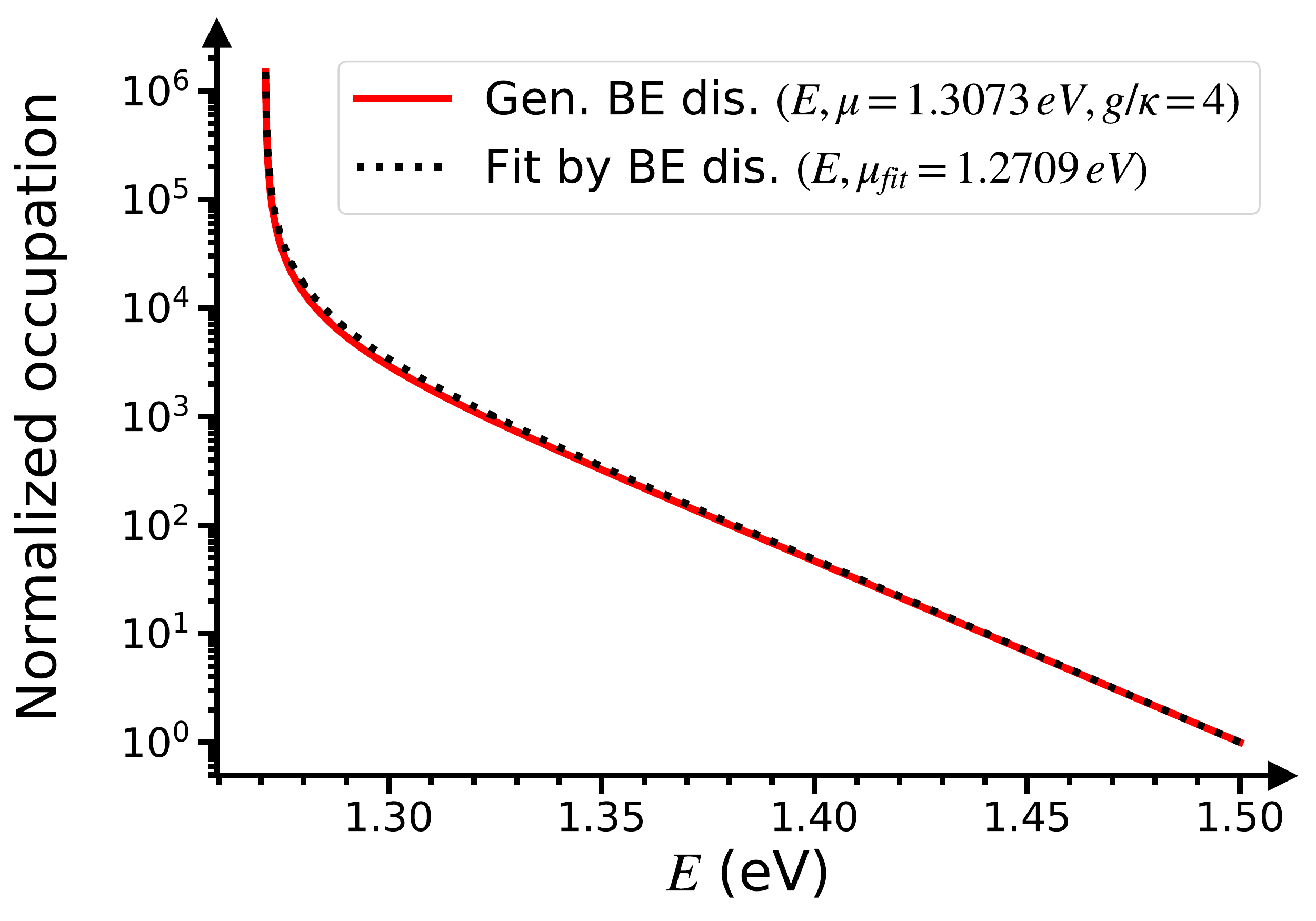}    
    \caption{Best fit of a generalized BE distribution ( Eq. (\ref{eq:photon_number_BE_laser_approach})) by an ideal BE distribution ( Eq. (\ref{eq:photon_number_BE_blackbody})). The fit parameter of the ideal BE distribution is the quasi-Fermi levels splitting $\mu_{fit}$. $g^l$ and $\kappa^l$ are assumed constant over all modes with $g/\kappa= 4$. All curves are normalized to 1 at $E= 1.50$ eV. Other parameters are compiled in Appendix \ref{app:parameters}.
    }
    \label{fig:BE_vs_gen_BE}
\end{figure}

\section{Model parameters values in figures}\label{app:parameters}

We take parameter values representative of the VCSEL used in Ref. \cite{barland}. The semiconductor gain medium consists of InGaAs quantum wells. We take $E^{gap}= 1.215$ eV, $m^*_c = 0.059\times m_e$, $m^*_v = 0.37\times m_e$ where $m_e$ is the electron mass \cite{coldren_corzine}. The hole mass corresponds to the valence band heavy-hole mass. Contribution of transitions with other valence bands is neglected. We assume room temperature operation $T=300$ K.

\section{Full expression of $g^{(2)}(0)$}\label{app:g_2_complet}

According to the treatment and notations of Sec. \ref{sec:modele_langevin}, the full expression of the second-order intensity correlations at zero-time
delay writes \cite{coldren_corzine, mork_lippi_2018,fedyanin}:

\begin{equation}\label{eq:}
\begin{split}
   & g^{(2)} (0)  = 1 - \frac{1}{N_{ss}} \\
   & + \frac{\gamma_{pe}^2 S_{ee} + \gamma_{pe} \gamma_{ee} (2S_{p\,e}) + (\gamma_{pe}\gamma_{ep} + \gamma_{pp}\gamma_{ee} + \gamma_{ee}^2)S_{pp}}{(\gamma_{pp} + \gamma_{ee})(\gamma_{pe}\gamma_{ep} + \gamma_{pp}\gamma_{ee}) N_{ss}^2
    },
\end{split}
\end{equation}

where $2S_{ee} = 2 S_{pp} = 2 R_{em}[N_{ss} +1]$ and $2S_{ep}= - R_{em}[2 N_{ss} +1] + N_{ss} [ R_{em} - R_{abs}]$.

This expression can be expanded more explicitely as 

\begin{equation}\label{eq:g2_fedyanin_simp}
\begin{split}
     & g^{(2)} (0)  =  1 - \frac{1}{N_{ss}}  + \Bigg[ \frac{1}{1 +  N_{ss} / N_{\beta}+ N_{ss}^2 / N_{CO}^2}  \Bigg] \\
     & \times \Bigg[ \frac{1}{1 +  N_{ss}^2 / N_{LAS}^2}  \Bigg] 
     \times \Bigg[ 1 + \frac{N_{ss}^2}{N_{LAS}^2}  +  \frac{N_{ss}}{N_{\beta}}  \\
     &  + \Big( \frac{N_{ss} }{N_{CO}} \Big)^2 \Big( \frac{\partial_{N_e} R_{em}}{\partial_{N_e} [R_{em} - R_{abs}] } + \frac{ N_{ss} }{ N_{LAS}^2 } + \kappa \frac{R_{em}}{g} \Big)
    \Bigg]
\end{split}
\end{equation}

where $N_{\beta} \approx \Big[ \frac{1}{\beta} - 1 \Big]^{-1} \frac{R_{em}(N_{ss}= \infty)}{ \partial_{N_e} R_{em} (N_{ss}= \infty) }$ and $N_{CO}\approx \bigg[ \frac{R_{em}(N_{ss}= \infty)}{ \partial_{N_e} [R_{em} - R_{abs}] (N_{ss}= \infty)} \bigg]^{1/2}$. In the limit $\beta \to 1$ (resp. $\beta \to 0$), the first (resp. second) term between brackets dominates while the product of the other terms $\approx 1$. The term $1/N_{ss}$ is always negligible.

\bibliography{ref}

\end{document}